\documentstyle[twocolumn,pra,aps,amssymb,psfig,amstex]{revtex}

\newcommand{\pfl}[1]{\underleftarrow{#1_{}}}
\newcommand{\pfr}[1]{\underrightarrow{#1_{}}}

\begin{document}

\draft

\title{3D input-output relations of quantized light \\ at dispersing
and absorbing multilayer dielectric plates}
\author{S.~Scheel and D.-G.~Welsch}
\address{Theoretisch-Physikalisches Institut, 
Friedrich-Schiller-Universit\"at Jena, Max-Wien-Platz 1, D-07743 Jena,
Germany}

\date{February 14, 2001}
\maketitle

\begin{abstract}
The theory developed by Gruner and Welsch
[Phys. Rev. A {\bf 54}, 1661 (1996)] for calculating the
1D input-output relations of the quantized electromagnetic field
at dispersing and absorbing dielectric (multilayer) plates is
generalized to the three-dimensional case.
First a general recipe for the derivation of the
reflection and transmission coefficients at an arbitrary
body that separates two half spaces from each other is presented.
The general theory is then applied to the
case of planar multilayer structures, for which the Green
tensor is well-known. 
\end{abstract}

%%%%%%%%%%%%%%%%%%%%%%%%%%%%%%%%%%%%%%%%%%%%%%%%%%%%%%%%%%%%%%%%%%%%%%
\section{Introduction}
\label{sec1}

Dielectric bodies such as beam splitters are vital ingredients
in quantum-optical experiments. Their properties are of utmost
importance for experiments with nonclassical light. For example,
losses due to absorption are known to degrade entanglement
\cite{Scheel00c}. The action on light of macroscopic bodies is
commonly described in terms of input-output relations.
Recently, the 1D input-output relations of the quantized
electromagnetic field at a multilayer dielectric plate \cite{Gruner96}
have been used for constructing CP maps for light interacting
with a dispersing and absorbing beam splitter \cite{Scheel00c,Knoll99}.

In the present article we generalize the
formalism given in \cite{Gruner96} and derive
the 3D input-output relations of the quantized
electromagnetic field at
a dispersing and absorbing multilayer dielectric plate.
The theory is essentially based on the
quantization of the macroscopic Maxwell field in dispersing
and absorbing dielectrics (for a review on this topic, see
Ref.~\cite{Buch}) where absorption is taken into account by
introducing (for each frequency $\omega$) a bosonic operator noise
current density which in turn is related to some fundamental
(collective) bosonic excitations of the electromagnetic field and
the dielectric matter as
\begin{equation}
\label{1.1}
\hat{\bf j}({\bf r},\omega) = \omega \sqrt{\frac{\hbar\epsilon_0}{\pi} 
\,\epsilon''({\bf r},\omega)} \,\hat{\bf f}({\bf r},\omega),
\end{equation}
with $\epsilon''({\bf r},\omega)$ being the imaginary part of the
complex permittivity function $\epsilon({\bf r},\omega)$]. Solving the 
corresponding inhomogeneous Helmholtz equation for
the electric-field operator in
terms of the classical Green tensor
${\sf G}({\bf r},{\bf s},\omega)$, one obtains
\begin{equation}
\label{1.2}
\hat{\bf E}({\bf r},\omega) = i\omega \mu_0 \int d^3{\bf s} \,
{\sf G}({\bf r},{\bf s},\omega) \,\hat{\bf j}({\bf s},\omega)
\end{equation}
as the representation of the (Schr\"odinger-) operator of the
electric-field strength in the frequency domain.

Let us assume that the (classical) Green tensor
of the electromagnetic field in the presence of the
dielectric body under study is known. The input-output
relations can then be derived by identifying the
contributions to the electromagnetic field in the
two half spaces with the input and output fields
and the (noise) field produced by the absorbing body. 
In Section \ref{sec2} we give the general recipe, which
we specify in Section \ref{sec3} for a multilayer
dielectric plate.

%%%%%%%%%%%%%%%%%%%%%%%%%%%%%%%%%%%%%%%%%%%%%%%%%%%%%%%%%%%%%%%%%%%%%%
\section{Identification of field parts}
\label{sec2}

\begin{figure}[h]
\hspace{1cm}
\psfig{file=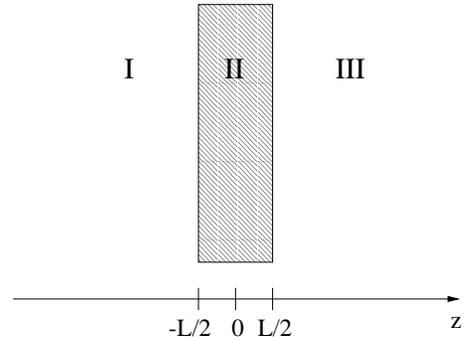,width=6cm}\\
\caption{\label{slab} Scheme depicting the dielectric slab located in
region II (\mbox{$-L/2$ $\!\le$ $\!z$ $\!\le$ $\!L/2$})
surrounded by vacuum in regions I and III.}
\end{figure}

Let us assume that the space ${\mathbb R}^3$ can be subdivided into
three separated regions I, II, and III, where the
region II represents a dielectric body. Although we
shall restrict our attention to an infinitely extended
plate of thickness $L$, the general results of this section 
remain also valid for other shapes of the body. Note that the assumption
of infinite extension of the plate can be justified by the
observation that, typically, the
cross-sectional area of an impinging light beam does not exceed the
surface area of the plate. 
For the sake of definiteness, let us denote the regions outside the
body with I (\mbox{$z$ $\!<$ $\!-L/2$}) and III
(\mbox{$z$ $\!>$ $\!L/2$}) (see Fig.~\ref{slab}).
\noindent

The electric-field operator (\ref{1.2})
in region I is given by 
\begin{equation}
\label{2.1}
\hat{\bf E}^{({\rm I})}({\bf r},\omega) = i\omega \mu_0
\int_{{\mathbb R}^3} d^3{\bf s} \,
{\sf G}^{({\rm I})}({\bf r},{\bf s},\omega) \, \hat{\bf j}({\bf s},\omega),
\end{equation}
where the Green tensor can be split into four parts \cite{Li94}:
\begin{eqnarray}
\label{2.2}
{\sf G}^{({\rm I})}({\bf r},{\bf s},\omega) &=&
{\sf G}^{(10)}({\bf r},{\bf s},\omega)
                               \hspace{2.1ex}\qquad ({\bf s}\in {\rm I})
\nonumber \\
&& + \,{\sf G}^{(11)}({\bf r},{\bf s},\omega) \qquad ({\bf s}\in {\rm I})
\nonumber \\
&& + \,{\sf G}^{(12)}({\bf r},{\bf s},\omega) \qquad ({\bf s}\in {\rm II})
\nonumber \\
&& + \,{\sf G}^{(13)}({\bf r},{\bf s},\omega) \qquad ({\bf s}\in {\rm III}).
\end{eqnarray}
Here, ${\sf G}^{(10)}({\bf r},{\bf s},\omega)$ is the solution
of the inhomogeneous Helmholtz equation (for the case where
region I extends the whole space to infinity), and the
${\sf G}^{(1i)}({\bf r},{\bf s},\omega)$ are solutions of the
homogeneous Helmholtz equation which insure the correct boundary
conditions at the surfaces of discontinuity. The electric-field
operator $\hat{\bf E}^{({\rm III})}({\bf r},\omega)$ in region III
is given accordingly. 

Let us consider the left
surface. The identification of the contributions of input
and output fields in Eq.~(\ref{2.1}) [together with Eq.~(\ref{2.2})]
is unique. The term determined by the free
Green tensor ${\sf G}^{(10)}({\bf r},{\bf s},\omega)$
represents the input field. The term determined by
${\sf G}^{(11)}({\bf r},{\bf s},\omega)$ is the contribution of the
field which is reflected at the surface, whereas the term
determined by ${\sf G}^{(13)}({\bf r},{\bf s},\omega)$ describes
the transmitted field through the dielectric body from
the right (region III). The reflected and transmitted fields
can then be combined to the output field. It is worth
noting that, since the Green tensor contains the full
information about the allowed electromagnetic-field structure,
the output field may also contain surface-guided waves
that propagate along the surface. In particular, surface-guided waves
can appear if the permittivity of the material satisfies the conditions
\mbox{$\epsilon'({\bf s},\omega)$ $\!<$ $\!0$} and
\mbox{$|\epsilon'({\bf s},\omega)|$ $\!\ll$
$\!|\epsilon''({\bf s},\omega)|$}.

{F}rom the above, we can subdivide the field at the left surface
(notation \mbox{${\bf r}|_{z=-L/2}$ $\!\equiv$ $\!{\bf r}^-_\perp$})
into the input field and the output field as
\begin{equation}
\label{2.3}
\hat{\bf E}_{\rm I}({\bf r}^-_\perp,\omega) =
\hat{\pfr{\bf E}}^{({\rm in})}_{\rm I}({\bf r}^-_\perp,\omega)
+\hat{\pfl{\bf E}}^{({\rm out})}_{\rm I}({\bf r}^-_\perp,\omega),
\end{equation}
with
\begin{equation}
\label{2.5}
\hat{\pfr{\bf E}}^{({\rm in})}_{\rm I}({\bf r}^-_\perp,\omega)
= i\omega \mu_0 \int_{\rm I} d^3{\bf s} \,
{\sf G}^{(10)}({\bf r}^-_\perp,{\bf s},\omega) \,
\hat{\bf j}({\bf s},\omega),
\end{equation}
\begin{eqnarray}
\label{2.6}
\lefteqn{
\hat{\pfl{\bf E}}^{({\rm out})}_{\rm I}({\bf r}^-_\perp,\omega)
= i\omega \mu_0 \int_{\rm I} d^3{\bf s} \,
{\sf G}^{(11)}({\bf r}^-_\perp,{\bf s},\omega)
\,\hat{\bf j}({\bf s},\omega)
}
\nonumber \\ && \hspace{10ex}
+\,i\omega \mu_0 \int_{\rm II} d^3{\bf s} \,
{\sf G}^{(12)}({\bf r}^-_\perp,{\bf s},\omega)
\,\hat{\bf j}({\bf s},\omega)
\nonumber \\ && \hspace{10ex}
+\,i\omega \mu_0 \int_{\rm III} d^3{\bf s} \,
{\sf G}^{(13)}({\bf r}^-_\perp,{\bf s},\omega)
\,\hat{\bf j}({\bf s},\omega).
\end{eqnarray}
Analogously, for the field at the right surface
(notation \mbox{${\bf r}|_{z=L/2}$ $\!\equiv$ $\!{\bf r}^+_\perp$})
we have
\begin{equation}
\label{2.7}
\hat{\bf E}_{\rm III}({\bf r}^+_\perp,\omega) =
\hat{\pfl{\bf E}}^{({\rm in})}_{\rm III}({\bf r}^+_\perp,\omega)
+\hat{\pfr{\bf E}}^{({\rm out})}_{\rm III}({\bf r}^+_\perp,\omega),
\end{equation}
with
\begin{equation}
\label{2.8}
\hat{\pfl{\bf E}}^{({\rm in})}_{\rm III}({\bf r}^+_\perp,\omega)
= i\omega \mu_0 \int_{\rm III} d^3{\bf s} \,
{\sf G}^{(30)}({\bf r}^+_\perp,{\bf s},\omega) \,
\hat{\bf j}({\bf s},\omega),
\end{equation}
\begin{eqnarray}
\label{2.9}
\lefteqn{
\hat{\pfr{\bf E}}^{({\rm out})}_{\rm III}({\bf r}^+_\perp,\omega)
= i\omega \mu_0 \int_{\rm I} d^3{\bf s} \,
{\sf G}^{(31)}({\bf r}^+_\perp,{\bf s},\omega)
\,\hat{\bf j}({\bf s},\omega)
}
\nonumber \\ && \hspace{10ex}
+\,i\omega \mu_0 \int_{\rm II} d^3{\bf s}
\,{\sf G}^{(32)}({\bf r}^+_\perp,{\bf s},\omega)
\,\hat{\bf j}({\bf s},\omega)
\nonumber \\ && \hspace{10ex}
+\,i\omega \mu_0 \int_{\rm III} d^3{\bf s}
{\sf G}^{(33)}({\bf r}^+_\perp,{\bf s},\omega)
\,\hat{\bf j}({\bf s},\omega).
\end{eqnarray}
The arrows indicate the propagation direction of the respective field
parts to larger ($\rightarrow$) or smaller ($\leftarrow$)
values of $z$. Recall that the output fields may also contain
surface-guided waves whose extensions
in the directions indicated by the arrows are small.

The first terms in Eqs.~(\ref{2.6}) and (\ref{2.9}) are the reflected
fields in the respective regions, the last terms are the transmitted
fields from the opposite sides of the body, whereas the second terms
arise from sources inside the body.
In order to derive the input-output relations, we have to rewrite
Eqs.~(\ref{2.6}) and (\ref{2.9}) in terms of the input fields at
both sides of the body and noise sources from inside the body.
This is always possible in a linear theory considered here, because
the superposition principle holds. Thus, we can
(formally) write 
\begin{eqnarray}
\label{2.12}
\lefteqn{
\hspace*{-4ex}
\hat{\pfl{\bf E}}^{({\rm out})}_{\rm I}
({\bf r}^-_\perp,\omega) =
\int d^2{\bf s}^-_\perp \,
{\sf R}_{\rm I}({\bf r}^-_\perp,{\bf s}^-_\perp,\omega)
\hat{\pfr{\bf E}}^{({\rm in})}_{\rm I}
({\bf s}^-_\perp,\omega)
}
\nonumber \\ && \hspace{6ex}
+\int d^2{\bf s}^+_\perp \,
{\sf T}_{\rm I,III}({\bf r}^-_\perp,{\bf s}^+_\perp,\omega)
\hat{\pfl{\bf E}}^{({\rm in})}_{\rm III}
({\bf s}^+_\perp,\omega)
\nonumber \\ && \hspace{6ex}
+\,\hat{\bf G}_{{\rm I},1}({\bf r}^-_\perp,\omega)
+\hat{\bf G}_{{\rm I},2}({\bf r}^-_\perp,\omega),
\end{eqnarray}
\begin{eqnarray}
\label{2.13}
\lefteqn{
\hspace*{-4ex}
\hat{\pfr{\bf E}}^{({\rm out})}_{\rm III}
({\bf r}^+_\perp,\omega)
= \int d^2{\bf s}^+_\perp \,
{\sf R}_{\rm III}({\bf r}^+_\perp,{\bf s}^+_\perp,\omega)
\hat{\pfl{\bf E}}^{({\rm in})}_{\rm III}
({\bf s}^+_\perp,\omega)
}
\nonumber \\ && \hspace{6ex}
+\int d^2{\bf s}^-_\perp \,
{\sf T}_{\rm III,I}({\bf r}^+_\perp,{\bf s}^-_\perp,\omega)
\hat{\pfr{\bf E}}^{({\rm in})}_{\rm I}
({\bf s}^-_\perp,\omega)
\nonumber \\ && \hspace{6ex}
+\,\hat{\bf G}_{{\rm III},1}({\bf r}^+_\perp,\omega)
+\hat{\bf G}_{{\rm III},2}({\bf r}^+_\perp,\omega),
\end{eqnarray}
with reflection coefficients ${\sf R}$ and transmission coefficients
${\sf T}$ (actually second-rank tensors). 
The integrations run over the respective surfaces of the body. For
example, the field at a point ${\bf r}^-_\perp$ is created by the
input field from the left hitting the surface at points
${\bf s}^-_\perp$ leading to the first term in Eq.~(\ref{2.12}), and
by the input field from the right hitting the surface at points
${\bf s}^+_\perp$ thus producing the second term in Eq.~(\ref{2.12}).
As already mentioned, surface-guided modes are generically included
in the formalism. The operators $\hat{\bf G}_{{\rm (I,III)},(1,2)}$
are related to the left and right propagating noise excitations 
inside the body.

The reflection and the transmission coefficients in general
depend on the observation point where the field is computed.
For example, ${\sf R}_{\rm I}({\bf r}^-_\perp,{\bf s}^-_\perp,\omega)$
obeys the Fredholm integral equation of the first kind
\begin{equation}
\label{2.14}
\int d^2{\bf s}^-_\perp \,
{\sf R}_{\rm I}({\bf r}^-_\perp,{\bf s}^-_\perp,\omega) \,
{\sf G}^{(10)}({\bf s}^-_\perp,{\bf s},\omega) \,
={\sf G}^{(11)}({\bf r^-_\perp},{\bf s},\omega),
\end{equation}
which has to be inverted to find 
${\sf R}_{\rm I}({\bf r}^-_\perp,{\bf s}^-_\perp,\omega)$.
Post-multiplying Eq.~(\ref{2.14}) by
$[{\sf G}^{(10)}({\bf s},{\bf x}^-_\perp,\omega)]^{-1}$,
integrating over ${\bf s}$, using the relation
\begin{equation}
\label{2.17}
\int_{\rm I} d^3{\bf s} \,
{\sf G}^{(10)}({\bf r}^-_\perp,{\bf s},\omega)
\left[ {\sf G}^{(10)}({\bf s},{\bf x}^-_\perp,\omega)
\right]^{-1} = \delta({\bf r}^-_\perp-{\bf x}^-_\perp),
\end{equation}
and renaming ${\bf x}^-_\perp$ as
${\bf s}^-_\perp$, we obtain the reflection coefficient as
\begin{eqnarray}
\label{2.18}
\lefteqn{
{\sf R}_{\rm I}({\bf r}^-_\perp,{\bf s}^-_\perp,\omega) 
}
\nonumber \\[.5ex] &&\hspace{4ex}
= \int_{\rm I} d^3{\bf s}\, {\sf G}^{(11)}({\bf r}^-_\perp,{\bf s},\omega)
\left[ {\sf G}^{(10)}({\bf s},{\bf s}^-_\perp,\omega)
\right]^{-1}.
\end{eqnarray}
Similarly, the transmission coefficient reads
\begin{eqnarray}
\label{2.19}
\lefteqn{
{\sf T}_{\rm I,III}({\bf r}^-_\perp,{\bf s}^+_\perp,\omega) 
}
\nonumber \\[.5ex] &&\hspace{4ex}
=\int_{\rm III} d^3{\bf s}\, {\sf G}^{(13)}({\bf r}^-_\perp,{\bf s},\omega)
\left[ {\sf G}^{(30)}({\bf s},{\bf s}^+_\perp,\omega)
\right]^{-1}.
\end{eqnarray}
Analogously, the reflection and transmission coefficients for
region III are derived to be 
\begin{eqnarray}
\label{2.20}
\lefteqn{
{\sf R}_{\rm III}({\bf r}^+_\perp,{\bf s}^+_\perp,\omega) 
}
\nonumber \\[.5ex] &&\hspace{4ex}
= \int_{\rm III} d^3{\bf s}\, {\sf G}^{(33)}({\bf r}^+_\perp,{\bf s},\omega)
\left[ {\sf G}^{(30)}({\bf s},{\bf s}^+_\perp,\omega)
\right]^{-1},
\end{eqnarray}
\begin{eqnarray}
\label{2.21}
\lefteqn{
{\sf T}_{\rm III,I}({\bf r}^+_\perp,{\bf s}^-_\perp,\omega)
}
\nonumber \\[.5ex] &&\hspace{4ex}
=\int_{\rm I} d^3{\bf s}\, {\sf G}^{(31)}({\bf r}^+_\perp,{\bf s},\omega)
\left[ {\sf G}^{(30)}({\bf s},{\bf s}^-_\perp,\omega)
\right]^{-1}.
\end{eqnarray}
Note, that the reflection and transmission coefficients are
polarization-dependent.
The remaining task is to invert the free Green tensors
${\sf G}^{(10)}({\bf s},{\bf s}^\pm_\perp,\omega)$ and
${\sf G}^{(30)}({\bf s},{\bf s}^\pm_\perp,\omega)$.
This can be done by expanding the inverse  Green tensors in terms of
a complete set of orthogonal solutions of the Helmholtz equation,
i.e. the TE and TM vector potentials, and using their respective
orthogonality relations.

%%%%%%%%%%%%%%%%%%%%%%%%%%%%%%%%%%%%%%%%%%%%%%%%%%%%%%%%%%%%%%%%%%%%%%
\section{Multilayer dielectric plates}
\label{sec3}

Beam splitters and related (passive) optical elements typically
consist of layers with different dielectric properties (for example,
anti-reflection coatings). Let us apply the formalism
developed in Section \ref{sec2} to the calculation of
the input-output relations of light at a multilayer dielectric
plate. To do so, we have to specify the Green tensor.
The Green tensor for planar multilayers can be found in
\cite{Li94,Chew,Tai} and is presented in the Appendix. All terms
contributing to the Green tensor can be given in terms of TE and TM
vector potentials thus making the distinction between different
polarizations very easy.

Suppose the plate consists of $N$ dielectric layers (hence,
region II is subdivided into $N$ subregions II$i$, \mbox{$i$ $\!=$
$\!1\ldots N$}). Then, the input fields are 
given by Eqs.~(\ref{2.5}) and (\ref{2.8}), and
the output fields are given by (\ref{2.6}) and (\ref{2.9})
with the identifications
\begin{eqnarray}
\int_{\rm II} d^3{\bf s} \,
{\sf G}^{(12)}({\bf r}^-_\perp,{\bf s},\omega) &\equiv& \sum\limits_{i=1}^N 
\int_{{\rm II}i} d^3{\bf s} \,
{\sf G}^{(12i)}({\bf r}^-_\perp,{\bf s},\omega),\\
\int_{\rm II} d^3{\bf s} \,
{\sf G}^{(32)}({\bf r}^+_\perp,{\bf s},\omega) &\equiv& \sum\limits_{i=1}^N 
\int_{{\rm II}i} d^3{\bf s} \,
{\sf G}^{(32i)}({\bf r}^+_\perp,{\bf s},\omega),
\end{eqnarray}
leading to the reflection and transmission
coefficients according to Eqs.~(\ref{2.18}) -- (\ref{2.21}), with
the Green tensor being specified in the Appendix. Note, that for planar
multilayers different polarizations do not mix, as it can be seen from 
the structure of the Green tensor, which contains only dyadic
products of vector functions of one type. This behaviour is
typical of planar layers.

{F}rom the Green tensor as given in the Appendix, it is not
difficult to identify the noise terms in Eqs.~(\ref{2.12}) and
(\ref{2.13}). For propagating waves it is the subdivision into left
and right moving waves in the slab. We derive
\begin{eqnarray}
\label{3.3}
\lefteqn{
\hat{\bf G}_{{\rm I},1}({\bf r}^-_\perp,\omega) =
-\frac{\omega\mu_0}{4\pi} \sum\limits_{i=1}^N
\int\limits_{{\rm II}i} \!d^3{\bf s}
\int_0^\infty \!d\lambda
\sum\limits_{n=0}^\infty
\bigg\{
\frac{2-\delta_{0n}}{\lambda h_{2i}}
}
\nonumber \\ &&\hspace{2ex}\times
\left[ A_M^{12i} {\bf M}_{{e\atop o}n\lambda}({\bf r}^-_\perp,h_1)
{\bf M}_{{e\atop o}n\lambda}({\bf s},-h_{2i})
\right.
\nonumber \\ &&\hspace{2ex}
\left. 
+\,A_N^{12i} {\bf N}_{{e\atop o}n\lambda}({\bf r}^-_\perp,h_1)
{\bf N}_{{e\atop o}n\lambda}({\bf s},-h_{2i}) \right]
\,\hat{\bf j}({\bf s},\omega)
\bigg\},
\end{eqnarray}
\begin{eqnarray}
\label{3.4}
\lefteqn{
\hat{\bf G}_{{\rm I},2}({\bf r}^-_\perp,\omega) =
-\frac{\omega\mu_0}{4\pi} \sum\limits_{i=1}^N
\int\limits_{{\rm II}i}\! d^3{\bf s}
\int_0^\infty \!d\lambda
\sum\limits_{n=0}^\infty
\bigg\{
\frac{2-\delta_{0n}}{\lambda h_{2i}}
}
\nonumber \\ &&\hspace{2ex}\times
\left[ B_M^{12i} {\bf M}_{{e\atop o}n\lambda}({\bf r}^-_\perp,h_1)
{\bf M}_{{e\atop o}n\lambda}({\bf s},h_{2i})
\right.
\nonumber \\ &&\hspace{2ex}
\left.
+\,B_N^{12i} {\bf N}_{{e\atop o}n\lambda}({\bf r}^-_\perp,h_1)
{\bf N}_{{e\atop o}n\lambda}({\bf s},h_{2i}) \right]
\,\hat{\bf j}({\bf s},\omega)
\bigg\},
\end{eqnarray}
\begin{eqnarray}
\label{3.5}
\lefteqn{
\hat{\bf G}_{{\rm III},1}({\bf r}^+_\perp,\omega) =
-\frac{\omega\mu_0}{4\pi} \sum\limits_{i=1}^N
\int\limits_{{\rm II}i} \!d^3{\bf s}
\int_0^\infty \!d\lambda
\sum\limits_{n=0}^\infty
\bigg\{
\frac{2-\delta_{0n}}{\lambda h_{2i}}
}
\nonumber \\ &&\hspace{2ex}\times
\left[ C_M^{32i} {\bf M}_{{e\atop o}n\lambda}({\bf r}^+_\perp,-h_3)
{\bf M}_{{e\atop o}n\lambda}({\bf s},-h_{2i})
\right.
\nonumber \\ &&\hspace{2ex}
\left.
+\,C_N^{32i} {\bf N}_{{e\atop o}n\lambda}({\bf r}^+_\perp,-h_3)
{\bf N}_{{e\atop o}n\lambda}({\bf s},-h_{2i}) \right]
\,\hat{\bf j}({\bf s},\omega)
\bigg\},
\end{eqnarray}
\begin{eqnarray}
\label{3.6}
\lefteqn{
\hat{\bf G}_{{\rm III},2}({\bf r}^+_\perp,\omega) =
-\frac{\omega\mu_0}{4\pi} \sum\limits_{i=1}^N
\int\limits_{{\rm II}i} \! d^3{\bf s}
\int_0^\infty \!d\lambda
\sum\limits_{n=0}^\infty
\bigg\{
\frac{2-\delta_{0n}}{\lambda h_{2i}}
} \nonumber \\ &&\hspace{2ex}\times
\left[ D_M^{32i} {\bf M}_{{e\atop o}n\lambda}({\bf r}^+_\perp,-h_3)
{\bf M}_{{e\atop o}n\lambda}({\bf s},h_{2i})
\right.
\nonumber \\ &&\hspace{2ex}
\left.
+\,D_N^{32i} {\bf N}_{{e\atop o}n\lambda}({\bf r}^+_\perp,-h_3)
{\bf N}_{{e\atop o}n\lambda}({\bf s},h_{2i}) \right]
\,\hat{\bf j}({\bf s},\omega)
\bigg\}.
\end{eqnarray}

Let us briefly compare the 3D input-output relations derived here with
1D input-output relations given in \cite{Gruner96}. Because of translational
invariance along the ($x,y$)-directions and the assumed horizontal
incident (propagation along the $z$-axis), the reflection and
transmission coefficients in \cite{Gruner96} do not depend on
spatial coordinates and  reduce to scalar functions (of frequency and
material parameters). Translational invariance along the
($x,y$)-directions is also assumed in the 3D treatment of
the problem, but here the reflection and transmission coefficients
must be regarded as space-dependent second-rank tensors in general.
Further, the output fields do not only contain
volume waves but also surface-guided waves.

%%%%%%%%%%%%%%%%%%%%%%%%%%%%%%%%%%%%%%%%%%%%%%%%%%%%%%%%%%%%%%%%%%%%%%
\section{Conclusions}
\label{sec4}

In this article we have developed a general concept for
deriving the 3D input-output relations of optical fields at
dispersing and absorbing bodies, and we have applied it to
a multilayer dielectric plate. Since the theory is based on
the quantization of the electromagnetic field in absorbing
media which relies on a source-quantity representation of the field
in terms of the (classical) Green tensor, the task reduces to the
determination of the Green tensor and its respective contributions
to the input and output fields.
In this way, the output fields can be related to the input
fields via transmission and reflection and to some noise
excitation associated with material absorption.  
Both volume waves and surface-guided waves are included
in the input-output relations.

%%%%%%%%%%%%%%%%%%%%%%%%%%%%%%%%%%%%%%%%%%%%%%%%%%%%%%%%%%%%%%%%%%%%%%
\appendix
\section{Green tensor for a dielectric layer}
\label{apA}

We briefly repeat some basic formulas for the Green tensor in the
spectral representation (for details, see \cite{Li94}).
The cylindrical TE and TM vector wave functions
for {\it e}ven and {\it o}dd waves which
are frequently used throughout are defined as
\begin{eqnarray}
\label{A1}
\lefteqn{
{\bf M}_{{e \atop o}n\lambda}({\bf r},h) =
\left[
\mp \frac{nJ_n(\lambda r)}{r} {\sin \choose \cos} n\psi \,{\bf e}_r
\right.
}
\nonumber\\[.5ex] &&\hspace{6ex}
\left.
-\,\frac{dJ_n(\lambda r)}{dr} {\cos \choose \sin} n\psi \,{\bf e}_\psi
\right] e^{ihz},
\end{eqnarray}
\begin{eqnarray}
\label{A2}
\lefteqn{
{\bf N}_{{e \atop o}n\lambda}({\bf r},h) =
\frac{1}{k} \left[
ih \frac{dJ_n(\lambda r)}{dr} {\cos \choose \sin} n\psi \,{\bf e}_r
\right.
}
\nonumber \\[.5ex] &&
\left.
\mp \frac{ihJ_n(\lambda r)}{r} {\sin \choose \cos} n\psi
\,{\bf e}_\psi
+\lambda^2 J_n(\lambda r) {\cos \choose \sin} n\psi \,
{\bf e}_z \right] e^{ihz}, \nonumber \\
\end{eqnarray}
where \mbox{$h$ $\!=$ $\!\sqrt{k^2-\lambda^2}$}, 
\mbox{$k^2$ $\!=$ $\!\epsilon(\omega)(\omega/c)^2$}, with
$\epsilon(\omega)$ being the complex permittivity
in the respective region of space.

The Green tensor at {\it s}ource point ${\bf s}$ and
{\it f}ield point ${\bf r}$ can always be decomposed into
a sum of an (unbounded) free Green tensor
${\sf G}^{(f0)}({\bf r},{\bf s},\omega)$ 
and the scattering Green tensor
${\sf G}^{(fs)}({\bf r},{\bf s},\omega)$ as
\begin{equation}
\label{A3}
{\sf G}^{(fs)}({\bf r},{\bf s},\omega) =
{\sf G}^{(f0)}({\bf r},{\bf s},\omega) \delta_{fs} +
{\sf G}^{(fs)}({\bf r},{\bf s},\omega) .
\end{equation}
Standard representations of the free Green tensor
${\sf G}^{(f0)}({\bf r},{\bf s},\omega)$ can be found, e.g., in
\cite{Buch,Li94,Chew,Tai}. The scattering Green tensor is
given in \cite{Li94} in the compact form of
\begin{eqnarray}
\label{A4}
\lefteqn{
{\sf G}^{(fs)}({\bf r},{\bf s},\omega) =
\frac{i}{4\pi} \int_0^\infty d\lambda
\sum\limits_{n=0}^\infty
\bigg\{
\frac{2-\delta_{0n}}{\lambda h_s}
}
\nonumber \\ &&\hspace{2ex}\times\,
\bigg[
(1-\delta_{f3}) {\bf M}_{{e\atop o}n\lambda}({\bf r},h_f)
\nonumber \\ &&\hspace{6ex}\times
\left[
(1\!-\!\delta_{s1}) A_M^{fs} {\bf M}_{{e \atop o}n\lambda}({\bf
s},-h_s)
\right.
\nonumber \\ &&\hspace{12ex}
\left.
+\,
(1\!-\!\delta_{s3}) B_M^{fs} {\bf M}_{{e \atop o}n\lambda}({\bf
s},h_s)
\right]
\nonumber \\&&\hspace{2ex} 
+\,(1-\delta_{f3}) {\bf N}_{{e\atop o}n\lambda}({\bf r},h_f) 
\nonumber \\ &&\hspace{6ex}\times
\left[
(1\!-\!\delta_{s1}) A_N^{fs} {\bf N}_{{e \atop o}n\lambda}({\bf
s},-h_s)
\right.
\nonumber \\ &&\hspace{12ex}
\left.
+\,
(1\!-\!\delta_{s3}) B_N^{fs} {\bf N}_{{e \atop o}n\lambda}({\bf
s},h_s)
\right]
\nonumber \\ &&\hspace{2ex}
+\,(1-\delta_{f1}) {\bf M}_{{e\atop o}n\lambda}({\bf r},-h_f) 
\nonumber \\ &&\hspace{6ex}\times
\left[
(1\!-\!\delta_{s1}) C_M^{fs} {\bf M}_{{e \atop o}n\lambda}({\bf
s},-h_s)
\right.
\nonumber \\ &&\hspace{12ex}
\left.
+\,
(1\!-\!\delta_{s3}) D_M^{fs} {\bf M}_{{e \atop o}n\lambda}({\bf
s},h_s)
\right]
\nonumber \\ &&\hspace{2ex}
+\,(1-\delta_{f1}) {\bf N}_{{e\atop o}n\lambda}({\bf r},-h_f) 
\nonumber \\ &&\hspace{6ex}\times
\left[
(1\!-\!\delta_{s1}) C_N^{fs} {\bf N}_{{e \atop o}n\lambda}({\bf
s},-h_s)
\right.
\nonumber\\&&\hspace{12ex}
\left.
+\,
(1\!-\!\delta_{s3}) D_N^{fs} {\bf N}_{{e \atop o}n\lambda}({\bf
s},h_s)
\right]
\bigg]\bigg\}, 
\end{eqnarray} 
where the scattering coefficients $A_{M,N}^{fs}$, $B_{M,N}^{fs}$,
$C_{M,N}^{fs}$, and $D_{M,N}^{fs}$ are determined by the boundary
conditions at the interfaces between the layers.
The notation used here is such that the indices of the {\it f}ield and 
{\it s}ource points cover the range ($1,2i,3$), $i$ $\!=$ $\!1\ldots N$,
distinguishing between region I (index $1$), region III (index $3$),
and $N$ layers II$i$ (indices $2i$) that build up region II.

%%%%%%%%%%%%%%%%%%%%%%%%%%%%%%%%%%%%%%%%%%%%%%%%%%%%%%%%%%%%%%%%%%%%%%

\end{document}